# Securing the Future Internet of Things with Post-Quantum Cryptography

Adarsh Kumar[1], Carlo Ottaviani[2], Sukhpal Singh Gill[3] and Rajkumar Buyya[4]

[1]Department of Systemics, School of Computer Science, University of Petroleum and Energy Studies, Dehradun, India
[2]Department of Computer Science & York Centre for Quantum Technologies, University of York, UK
[3]School of Electronic Engineering and Computer Science, Queen Mary University of London, UK
[4]Cloud Computing and Distributed Systems (CLOUDS) Laboratory, School of Computing and Information Systems, The University of Melbourne, Australia

Correspondence: Sukhpal Singh Gill, School of Electronic Engineering and Computer Science, Queen Mary University of London, Mile End Road, Bethnal Green, London E1 4NS, UK.
Email: s.s.gill@qmul.ac.uk

**Abstract**

**Traditional and lightweight cryptography primitives and protocols are insecure against quantum attacks. Thus, a real-time application using traditional or lightweight cryptography primitives and protocols does not ensure full-proof security. Post-quantum Cryptography is important for the Internet of Things (IoT) due to its security against Quantum attacks. This paper offers a broad literature analysis of post-quantum cryptography for IoT networks, including the challenges and research directions to adopt in real-time applications. The work draws focus towards post-quantum cryptosystems that are useful for resource-constraint devices. Further, those quantum attacks are surveyed, which may occur over traditional and lightweight cryptographic primitives.**

## 1. Introduction

Internet of Things (IoT)-based applications such as smart cities, healthcare, meteorology, agriculture, and smart grids usually make use of tiny and affordable resource-constrained devices. To secure these resource-constrained devices, lightweight security primitives and protocols are required. Quantum computers are expected to break traditional lightweight security primitives and protocols. Additionally, IoT networks are vulnerable to various attacks including Sybil, eclipse, replay, side-channel, and false data injection. Thus, lightweight post-quantum cryptography mechanisms are required to be integrated. Lightweight post-quantum cryptography mechanisms mainly include lattice-based cryptography, code-based cryptography, multivariate polynomial cryptography, hash-based signatures, and others. Figure 1 shows an overview of post-quantum cryptography and associated security terms and classifications.

This work aims to perform a short survey and analysis of the importance of IoT networks from a futuristic point of view. In futuristic IoT networks, the chances of quantum attacks and security using post-quantum cryptography are required to be analyzed. Further, the research and technical challenges in integrating post-quantum cryptography with IoT networks must be conducted to ensure the high standardization of security in futuristic IoT networks.

This rest of the sections are organized as follows. Section 2 presents the futuristic importance of the Internet of Things (IoT). Section 3 presents the major challenges and possible solutions in futuristic IoT networks. Section 4 shows the important quantum attacks observed in resource-constrained networks or with lightweight cryptography primitives and protocols. Section 5 describes the post-quantum cryptography types and recent developments. This section performs the comparative analysis of recent approaches in these areas as well. Section 6 presents the futuristic research directions in integrating post-quantum cryptography aspects with IoT applications to secure it from quantum attacks. Finally, the conclusion is drawn in section 7.

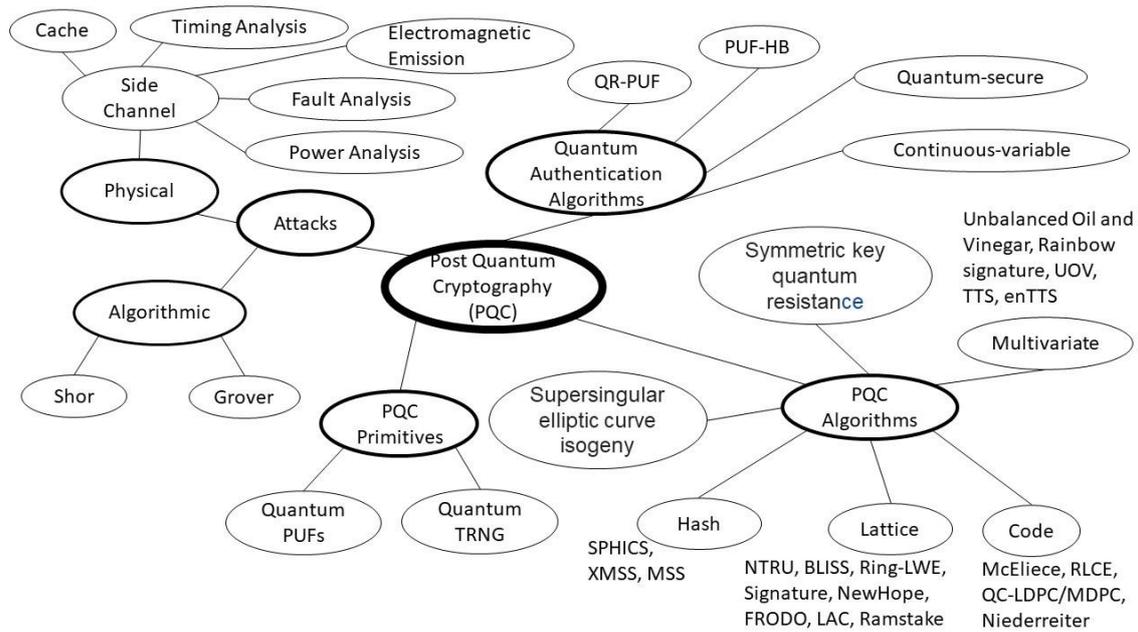

**Figure 1:** Post-quantum cryptography and its connected areas

## 2. Rising IoT

In the future, it is expected that more and more devices will be interconnected to the Internet compared to people joining the Internet services. With this growth, a large amount of data will be generated. For example, IoT-based applications like smart homes, smart traffic light systems, smart transportation systems, automated traffic lights, automated vehicles, medical and healthcare services, and other supply chain management will generate huge data per second. Major of these applications use small and resource-constrained devices (belong to Class 0, Class 1, or Class 2 devices with less than 10 KB of RAM or storage space and can have less than 100 KB of code in their flash memory). Class 0 contains extremely constrained resource devices, Class 1 is more powerful than Class 0 but still having very constrained resources, and Class 2 is more powerful than Class 0 and Class 1 but have sufficient power and memory available for required computational tasks. By 2024, It is expected that there will be more than 200 billion IoT devices (of all classes i.e. Class 0, Class 1, and Class 2) interconnected worldwide [1].

## 3. Challenges and possible solutions of future IoT networks

Various challenges in traditional resource-constrained IoT devices [2] are discussed as follows:

- Fast evolving quantum computing approaches are continuously put challenges to traditional cryptography enabled IoT applications. Although these challenges are mathematical and theoretical in the present time at large there is no guarantee that futuristic pre-quantum or post-quantum cryptography-based IoT applications will be capable of resisting quantum algorithms or attacks [2]. Thus, there is a strong need to focus on those approaches.
- Resource-constrained IoT networks use much smaller key sizes (usually 128 bits to 4096 bits). However, post-quantum algorithms require much larger key sizes. Thus, integration of IoT networks with post-quantum cryptography algorithms requires analysis of key size, security levels, network performances, and scalability.
- Resource-constrained devices induce latency during cryptography primitives and protocols integration at both source and destination ends. This makes it difficult for fast computing servers to synchronize communication with IoT devices. For instance, there is a limit on the number of signatures in the hash-based post-quantum cryptosystem. Thus, there is a need to generate a new set of keys for a subsequent group of messages that is not an easy job for traditional resource-constrained-based IoT devices. It will consume large energy and will not give efficient results. Thus, either there is a need to redesign post-quantum approaches for resource-constrained IoT devices that efficiently handle network performances and Quality of Service (QoS).

- Public-key or digital signature post-quantum algorithms consume resources, energy and add delays. Thus, a lack of optimization approaches or high-accuracy measurement methods put major hurdles. Designing and developing such approaches/methods can help in selecting efficient approaches and discard others.
- The majority of post-quantum cryptosystems focus on proving high security by somehow neglecting the other parameters. Among other parameters, energy, delay, and resource consumption are important for IoT networks. An approach acceptable to real-time applications should include multiple parameters rather than a few.
- Presently, there is no standard approach to measure the security levels of post-quantum cryptosystems against quantum attacks. Thus, there is a need to identify the parameters and their priority levels for IoT applications. This is possible if some standards (like NIST, IEEE, IETF) design a set of procedures and processes, or leading researchers in this area work together and conclude it in some framework.
- In future resource-constrained IoT networks, devices are expected to be more powerful. In nearby times, these devices are expected to remain low-computational. However, the computational ability is expected to be increased from after 10 to 20 years. Among all scenarios, energy-efficient post-quantum approaches are expected to be useful for real scenarios provided these approaches do not compensate over security with time.
- Optimization of quantum algorithms for IoT nodes is required for IoT devices. For example, the lattice-based cryptosystem in the post-quantum cryptography world is a prime candidate for IoT devices. In this algorithm, there is a need to speed up the polynomial-based multiplication calculation and reduce the energy consumption and execution time. Likewise, there are a set of computational tasks in other post-quantum cryptography mechanisms (in a finite field) that need optimization for IoT devices.
- Lack of network architectures that help in understanding the distributed computing architecture is necessary. This architecture can help the assembler code for an IoT microcontroller to be optimized. There are various optimization approach that relies upon the architecture. For example, loop optimization, and register optimization techniques.

## 4. Quantum Attacks

In this section, we discuss quantum-attacks that can be implemented over traditional or lightweight cryptography approaches useful in IoT networks for ensuring confidentiality, integrity, authentication, availability, and non-repudiation [3].

### 4.1 Shor's Algorithm

This algorithm [5] is polynomially efficient in solving integer factorization, discrete logarithmic problem, and elliptic-curve-based discrete logarithmic problem. If 'n' bits is considered to be the key size used in these algorithms, then the solution to these approaches can be found out in $O(n^3)$. Shor algorithm is capable, in principle, of breaking any traditional asymmetric key-distribution algorithm. The lightweight cryptographic algorithms, suitable for IoT devices, apply less hard problems compared to the traditional approach. Thus, there is a need to find out solutions that use hard problems to tackle Shor's algorithm.

### 4.2 Grover's Algorithm

The discovery of the Grover algorithm showed that quantum computers have a quadratic speed-up in searching databases compared to classical computers. As the heart of the algorithm there is the use of quantum superposition in its design, optimal in applying parallel execution, and well-known for applying unstructured search with high probability of unique output.

**4.3 Side-Channel Attack**

In this attack, the eavesdropper tries to exploit the vulnerabilities in implementation or environment rather than mathematical structures. For example, an invasive side-channel attack can do anything with cryptographic devices whereas, a non-invasive side-channel attack does not physically tamper with the device but harms using timing attacks, power analysis, electromagnetic attacks, etc. In a semi-invasive side-channel attack, the goal of an attacker is to add fault in the algorithms or cryptosystems. Here, an attacker tries to observe the system outcomes after maliciously inserting the faults which in turn may leak some useful information. This is a category of attack which is frequently discussed in the post-quantum cryptography world. Side-channel attacks represent a serious threat also for traditional quantum key distribution protocols, e.g., BB84-like schemes [22]. Partial mitigations to this threat can be obtained by the so called *device-independent Quantum key distribution (QKD)*, whose original principles can be track back to the seminal work of Ekert [23]. These class of key-distribution schemes allow a stronger security at the price of more limited performances in terms of speed and rate [24].

**4.4 Multi-target Pre-image Search Attack**

In 1994, van Oorschot-Wiener proposed "parallel rho method". This is a low computation algorithm with a parallel pre-quantum multi-target preimage search. This algorithm is a security challenge for symmetric cryptography approaches especially AES-128. This algorithm targets to find the AES keys by executing fast steps over a mesh of processors.

**4.5 Attack Strategies for Quantum Key Distribution**

Quantum cryptography promise of delivering a perfectly secure secret-key is valid only from an information theoretic view-point. In practical terms, QKD protocols are never perfect, and trade-offs exist on key-rate performances and security depending on the assumptions made to apply the security proofs, characterize the devices used in the realistic implementation. In addition to that, other parameters may affect security and performance: number of signal exchanged, postprocessing efficiency, level of noise affecting the signals exchanged.

Quantum cryptography can be divided in two families of protocols, characterized by the physical systems used to perform the encoding of the classical information: Discrete-variable (DV) and continuous-variable (CV) QKD protocols. In DV, which are the earliest being introduced (see e.g. the BB84 and E92 protocols), use qubits to encode information and the ideal, information-theoretic, security of this approach relies on the possibility of having a genuine source of single photons. CV protocols use instead more intense pulses, and their encoding is made using the amplitude and phases of intense pulses of coherent light. DV variable approach has its main advantage in that it is more suitable for long-distance communication, while CV may allow to achieve higher key-rates, but the range is by construction more sensible to the noise of the channel, and so more limited in term of achievable distances. Several attacks may be designed to eavesdrop both DV and CV-QKD.

For DV protocols, Photon-number-splitting (PNS) attacks are the first to consider: Realistic QKD implementations do not use truly single-photon sources, but rather strongly attenuated photon-sources, which will include more than one photon. That allows the eavesdropper (Eve) to deviate all photons in excess from the main quantum channel and store them in a quantum memory to be measured later, after all classical communication between the parties took place, to extract information. The receiver (Bob) and the sender, expecting single-photon pulses, will not notice any errors when they compare their bits during the sifting of the raw key. In such a way Eve may extract perfect information about the bits shared using multi-phonon pulses. However, the fact of having to employ multi-photon sources may be used to fight against PNS attack, via the use of decoy states where the sender (Alice) replaces, randomly, the multiphoton signal states with decoy states that Eve cannot distinguish. In such a way the parties can quantify the presence of the eavesdropper using a PNS attack and they can apply countermeasures like increasing the amount of error correction and privacy amplification or abort the protocol if the preparation of a key is not possible.

Another class of attacks are Trojan-Horse attack, where the eavesdropper attempts to acquire information by sending quantum signals within the parties' devices. For example, Eve may send pulses of light to the trusted devices (signal modulator, detectors) to gain information on the parties' measurement basis from the reflected lights. In some case knowing the measurement basis of the receiver may be sufficient to gain a complete knowledge of the key. Countermeasures exists to this class of attack and may depend on the specific Trojan-horse implementation. They may range from adding additional layer of devices to monitor outcoming and incoming photon from the trusted devices, extra attenuator, and phase randomization of sent qubits, to minimize Eve's achievable information.

Backflash attacks are a class of attack connected to the use of Avalanche Photodetectors, which may emit light when they detect. This emitted light (Backflash) may be collected by the eavesdropper and inform about components used by the parties, and the measurement basis adopted by the parties in each instance of the protocol. These kinds of attacks may be mitigated using spectral filters.

Class of attacks specific to CV protocols are those performed on the local oscillator (LO) and saturation attacks performed on the homodyne detectors typically used in this class of protocols. The LO is an intense reference laser pulse shared between the parties for synchronization of devises and calibration, in particular, of the vacuum shot-noise. To prevent attacks on the LO, it has been suggested to monitor the intensity of the LO, and, recently, it has also been suggested to employ local LO.

In saturation attacks the mismatch between security proof accuracy and assumption on realistic devices (the homodyne detection) is exploited. Homodyne detection saturates, above a certain level, this causes the receivers measurement to obtain measurement that may underestimate the noise introduced by the eavesdropper and so underestimating the amount of information gained by Eve. Counter measures have been described, relying on Gaussian filtering and post-selection to be sure that the signals used for the key are not falling above the level of saturation of the detector. Previous attacks may be grouped within the class of side-channel attacks [6] where the eavesdropper can exploit imperfections in the devices of the parties to gain information useful then extract, directly, or indirectly, information on the key-bits shared between Alice and Bob.

## 5. Post-Quantum Cryptography Types

Post-quantum cryptography is developed to resist quantum computers and quantum computing-based attacks. Various post-quantum cryptography approaches are already implemented for information and communication technologies. The major categories of post-quantum cryptography are hash, code, lattice, multivariate, and super-singular [10]. Figure 2 shows the algorithms developed in these categories and submitted to NIST for evaluation in 3 different rounds [4]. These categories and their importance to resource-constraint IoT networks are explained as follows [4],[7] [8] [9]. In NIST Post-quantum cryptography standardization, 23 signature and 59 encryption schemes were submitted in the initial round at the end of 2017. Out of these 82 schemes, 69 schemes were found to be complete and proper. After 3rd round, seven schemes were announced. However, NIST has to publish the standardization document by 2024, with the exception that some major breakthroughs happening in the quantum computing domain.

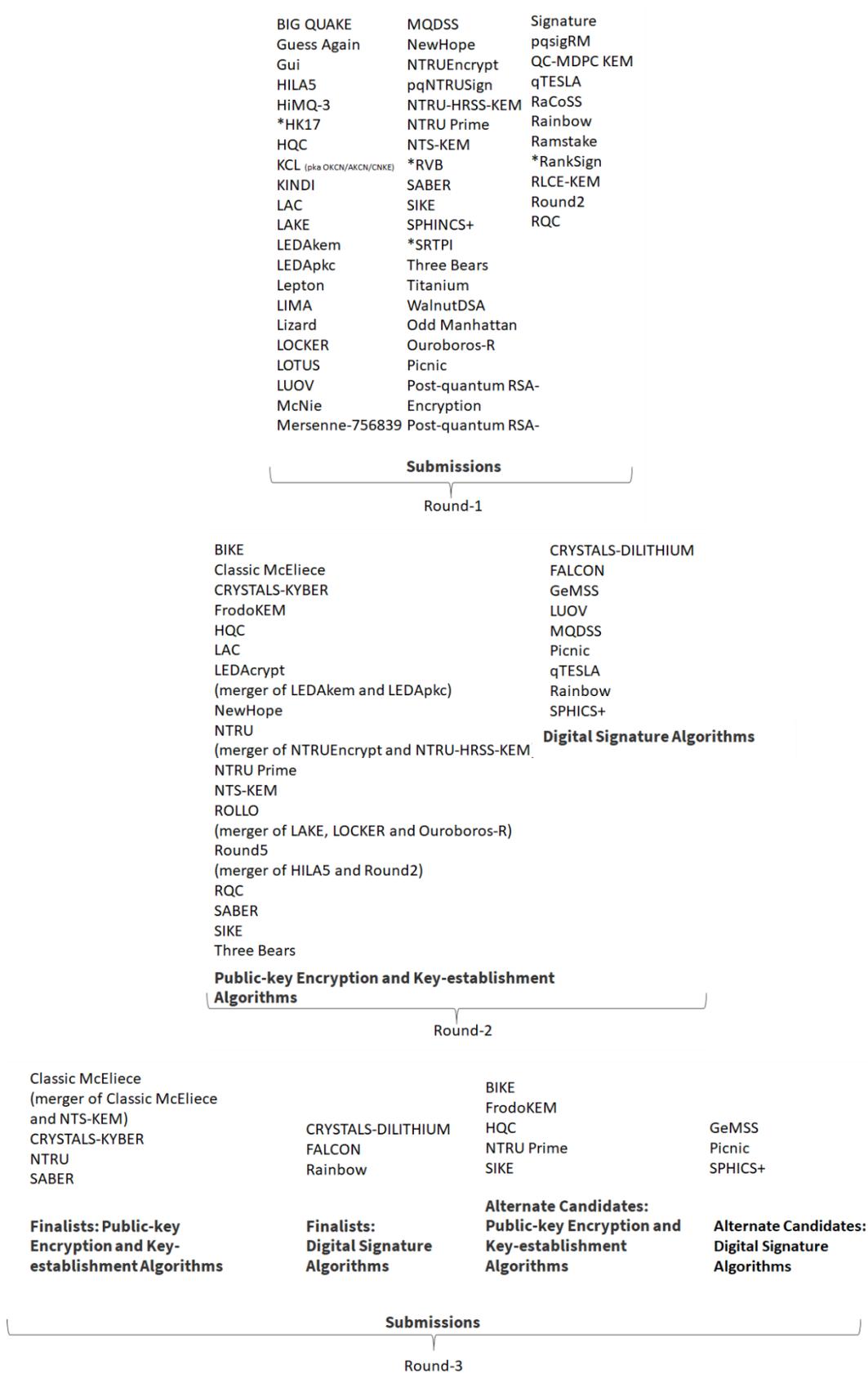

**Figure 2:** NIST Post-quantum Cryptography Algorithm Progresses in 3-Rounds

## 5.1 Lattice-based Cryptography

IoT-based companies are largely focusing on developing lightweight, cheaper and smarter products [9] without much caring about the security aspects. Quantum threats to cryptography in smart IoT devices-based networks, and services are also feasible. Therefore, post-quantum security aspects are required to be considered while designing the security solutions for smart IoT devices. The strong security, wide applicability, and efficiency to protect against attacks make the lattice-based cryptosystem the prime candidate for the future. The random key generation process is frequently used in various cryptosystems. This random key generation process requires average-case intractability or hard problems, and lattice-based cryptography has this feature [10]. The worst-case reduction to the average-case hard problem needs a selection of proper parameter that is easier in lattice-based cryptography for smart IoT devices. Since size (usually smaller) is required for smart IoT devices for fast operations. Lattice-based algorithms also prefer to use matrices and vectors in small order fields or rings with small size parameters. The uniqueness of this property between IoT and lattice-based algorithms makes them more suitable for security proposals. The challenge of proposing a security solution for IoT networks using lattice-based cryptosystem is the goal of interconnecting everything that requires lightweight cryptography primitives and protocols suitable for both resourceful and resource-constraint smart IoT devices. Examples of lattice-based cryptography include NTRUEncrypt, R-LWEenc, R-BIN-LWEenc, IBE, BLISS, and NewHope [9]. The bit security of these algorithms varies from 46 to 128 with the feasibility of their implementation over 8 to 32-bit CPU, and execution time varies from 1.9 to 317.4 msec.

## 5.2 Code-based Cryptography

McEliece cryptosystem is the first cryptosystem proposed in the asymmetric key encryption scheme that uses Goppa code and random generator matric using that code [11]. Code-based cryptosystems are considered to be not suitable for resource-constraint devices because of large memory requirements, large public key sizes, and long ciphertexts compared to lattice-based cryptography. Various code-based cryptography schemes are the McEliece cryptosystem and Niederreiter cryptosystem. Niederreiter cryptosystem is a variation of the McEliece cryptosystem and provides the same security. However, the encryption process in Niederreiter cryptosystem is ten times faster than the McEliece cryptosystem. Various attempts have been made to implement a lightweight solution for IoT devices [13]. Here, dRANKula, ROLLO, and Rank Quasi Cyclic (RQC) code-based schemes are explored to find the possibilities of implementation over resource-constraint devices.

## 5.3 Multivariate Polynomial Cryptography

Multivariate polynomial cryptography scheme uses simple arithmetic operations for security primitives and protocols. These operations include addition and multiplication in small finite fields. These simple operations and computations make this an important candidate for ensuring security in resource-constraint devices that include RFID cards, sensors, actuators, and smart cards. For example, the multivariate signature scheme gives a short signature that is of few hundred bits in length. As compared to other post-quantum cryptography schemes, the size of the signature is much shorter. The small size increases the speed of these mechanisms as well which in turn improves the performance, and makes this a promising candidate for IoT networks.

## 5.4 Hash-based Signatures

In [11], various features of the hash-based scheme are identified that are useful to the IoT ecosystem. Hash-based schemes rely solely on cryptographic hash functions rather than any other cryptographic assumptions such as number-theory-based hardness. Thus, it reduces the opportunities for cryptanalysis. This reduces the complexity of the overall system. The hash-based scheme is inherently dependent upon the application-specific environment which in turn enforces this scheme to be flexible in selecting the hash function to achieve the desired performance. Hash-functions have collision resistance, pre-image resistant, and second-pre-image resistant features that make this scheme protect the application against various attacks [13]. The hash-based scheme is having an option to protect IoT applications against those adversaries that steal the credentials from long-time running resource-constraint devices or mission-critical devices. Thus, a hash-based scheme creates a trustworthy and fair data-protected quantum IoT ecosystem. Lightweight versions of hash functions are available that open up an option for IoT applications to select resource-constraint device's suitable parameters that improve the network performance. One-way functionality of hash functions in hash-based scheme makes this scheme secure with backward and forward secrecies.

**5.5 Isogeny-based Cryptosystem**

This cryptosystem is based on supersingular elliptic curve isogenies [14]. This scheme can be used in a digital signature or key exchange approach. For example, the Supersingular Isogeny Diffie-Hellman Key Exchange (SIDH) approach is based on supersingular isogeny graphs and it is protected against cryptanalytic attacks from any adversary. SIDH provides the feasibility to have a small key size and 128-bit quantum security level. The features, including attack resistance and small size cryptography implementation, make this post-quantum cryptosystem a viable option for resource-constraint devices in IoT networks.

Table 1 performs a comparative analysis of recent work conducted over post-quantum cryptography. In this comparative analysis, it has been observed that security analysis, complexity computation, experimentation and deployment are important aspects to study in the post-quantum cryptography domain [24]. Further, a major focus in recent studies is drawn towards the analysis of possible designs in various post-quantum cryptography types.

**Table 1:** Comparative analysis of post-quantum cryptography

| Author | Year | A | B | C | D | E | F | G | H | I | Major Observations | Future Directions |
|---|---|---|---|---|---|---|---|---|---|---|---|---|
| Howe et al. [17] | 2021 | ✓ | ✓ | ✓ | ✓ | ✓ | ✓ | ✓ | ✗ | ✓ | This study has performed an analysis of post-quantum cryptography and its types. This survey includes analysis based on paradigm, implementation and deployment aspects. | This work can be extended to include the comparative security and complexity analysis of implementation and deployment aspects. |
| Bisheh-Niasar et al. [18] | 2021 | ✓ | ✗ | ✗ | ✗ | ✓ | ✓ | ✓ | ✓ | ✓ | In this work, optimization strategies are proposed in efficiency improvement and performance analysis. Further, an architecture is implemented for key exchange. | This work can be extended to include comparative analysis of optimization approaches or other optimization approaches like simulation annealing-based optimization that can be experimented with to improve efficiency and performance. |
| Fritzmann et al. [19] | 2021 | ✓ | ✓ | ✗ | ✓ | ✗ | ✓ | ✓ | ✓ | ✓ | This work has proposed masked hardware and software-based co-design for NIST post-quantum cryptography finalists Kyber and Saber. Further, a designed security and attack analysis is performed. | As claimed, most of the implemented algorithms are extendable for performing higher-order side-channel security. Similarly, other quantum, physical and classical attacks can be analyzed for the proposed masked-based approach. |
| Fritzmann et al. [20] | 2021 | ✓ | ✓ | ✗ | ✓ | ✗ | ✓ | ✗ | ✓ | ✗ | This work has conducted performance exploration of AURIX microcontroller for four lattice-based algorithms. Further, the security capabilities of ThreeBears are improved | This work can be extended to perform security analysis of proposed algorithms. Further, the complexity analysis can be extended with time and space complexity analysis. This |

| | | A | B | C | D | E | F | G | H | I | | |
|---|---|---|---|---|---|---|---|---|---|---|---|---|
| | | | | | | | | | | | using error-correction codes. | includes analysis of algorithm implementation, deployment and security prediction. |
| Cohen et al. [21] | 2021 | ✗ | ✓ | ✗ | ✗ | ✗ | ✓ | ✗ | ✓ | ✗ | This work has proposed a coding scheme to secure computational and information-theoretical aspects in the capacity of a network that uses a public key encryption scheme. Further, performance is analyzed to prove its suitability in the network. | The proposed public-key cryptography-based approach can be extended with hybrid solutions that combine information-theoretical security with public-key cryptography. The trade-off between security and information rate can further be analyzed for this hybrid scheme. |

A: Lattice-based Cryptography, B: Code-based Cryptography, C: Multivariate Polynomial Cryptography, D: Hash-based Signatures, E: Isogeny-based Cryptosystem, F: Security against quantum attacks, G: Complexity and performance analysis, H: Experimental analysis, I: Theoretical analysis.

## 6. Future Research Directions in Integrating Post-quantum Cryptography for IoT-applications

The important research directions identified in recent times to integrate post-quantum cryptography and/or other technologies for IoT applications are discussed as follows.

- Post-quantum cryptography uses various cryptographic primitives (hash, digital signature, symmetric and asymmetric keys, random number generations) that are common with blockchain-based technology. Integration of blockchain and post-quantum cryptography aspects can provide various security features including transparency, immutability, fault tolerance, resistance from quantum attacks, and adaptability to a large set of businesses. In [15][16], some architectures are proposed to integrate the two technologies (blockchain and quantum computing) for IoT networks. However, more technical solutions and evaluations are required to identify the best possible solution for resource constraint devices that are fast, energy-efficiency as well.
- In recent times [12] [15], it has been observed that few post-quantum cryptography standardization efforts and projects are started. To increase the adaptability of existing post-quantum cryptography schemes with IoT networks, there is a need to put more effort into those primitives and protocols that are not touched yet. Further, contributions towards this direction can be improved if the identification of parameters for standardization should be done in the early stages.
- Another major challenge is infrastructure requirements for post-quantum cryptography and IoT integrated ecosystem. To operate the post-quantum cryptography for IoT applications, there is a need to provide an environment that supports quantum computing, cost-effective hardware that supports quantum computing, terrestrial quantum networks, and adaptability analysis of existing standards with an integrated environment.
- IoT networks apply to a diverse set of applications and their issues such as performance issues in the smart city network. Resource-constraint devices-based sensor, RFID, or actuator-based network, data security issues for medical or healthcare applications, long-range communications for industrial IoT networks, and accessibility and traceability issues in supply chain management. The addition of quantum computing can speed up information processing in all scenarios, and post-quantum cryptography will fill up the necessary security requirements. Thus, exploring the quantum-IoT integrated environment for domain-specific issues would be interesting to explore in the future.
- Risk assessment in different domains like quantum risk assessment, cyber risk assessment, network failure risk assessments, etc. is important to analysis for post-quantum cryptography integrated IoT ecosystem. Continuous evaluations based on the risk assessment report during the lifecycle and deployment of IoT application is an interesting research direction. This needs to be explored with the advancement of technologies, and associated risks.

- The scarcity of quantum resources can increase the cost of implementation for IoT applications. Thus, there is a need to perform cost estimation for IoT applications. Here, there is a need for those experts that have experience in this domain to implement the quantum computing integrated platforms with standardized technologies. Additionally, there is a need to train or identify those experts as well who have the IoT infrastructure handling knowledge.
- IoT-based critical infrastructure requires higher interconnectivity and interdependencies. For example healthcare and public transportation networks. The increase in the number of users, and amount of data increases the threats to the system as well. Thus, there is a need to explore those threat models that can understand the requirements of life systems.

## 7. Conclusions

Recent studies have explored various post-quantum public-key cryptosystems for resource-constrained IoT devices. They identified lattice-based and hash-based schemes as prime candidates for IoT networks. All of these schemes are found to be as efficient and powerful as traditional cryptosystems. With the growth of interconnection of devices in IoT networks worldwide, lightweight and secure post-quantum cryptography approach for small devices with 32-bit architecture, 128-bit quantum security level, high-speed execution, and attack resistant features are expected to be developed in nearby times to fulfil the needs of future IoT networks.

## Data Availability Statement

Data sharing not applicable to this article as no datasets were generated or analysed during the current study.

## ORCID

**Adarsh Kumar** https://orcid.org/0000-0003-2919-6302

**Carlo Ottaviani** https://orcid.org/0000-0002-0032-3999

**Sukhpal Singh Gill** https://orcid.org/0000-0002-3913-0369

**Rajkumar Buyya** https://orcid.org/0000-0001-9754-6496